\documentclass[twocolumn,aps,prd,superscriptaddress,nofootinbib,showpacs]{revtex4}
\usepackage{graphicx}
\graphicspath{{figures/}{fig/}}
\usepackage{amsmath}
\usepackage{bm}
\usepackage{slashed}
\usepackage{epsfig}
\usepackage{amsfonts}
\usepackage{epstopdf}
\usepackage{hyperref}
\newcommand{\ben}{\begin{eqnarray}}
\newcommand{\een}{\end{eqnarray}}

\newcommand{\bef}{\begin{figure}[!htp]}
\newcommand{\eef}{\end{figure}}

\newcommand{\bea}{\begin{eqnarray}}
\newcommand{\eea}{\end{eqnarray}}

\def\ba{\begin{linenomath*}\begin{equation}}
\def\ea{\end{equation}\end{linenomath*}}

\begin{document}
\title{On the Renormalizability of Quasi Parton Distribution Functions}

\author{Tomomi Ishikawa}
\email{tomomi.ik@gmail.com}
\affiliation{T.~D.~Lee Institute, Shanghai Jiao Tong University,
Shanghai, 200240, P. R. China}
\author{Yan-Qing Ma}
\email{yqma@pku.edu.cn}
\affiliation{School of Physics and State Key Laboratory of Nuclear Physics and
Technology, Peking University, Beijing 100871, China}
\affiliation{Center for High Energy physics, Peking University, Beijing 100871, China}
\affiliation{Collaborative Innovation Center of Quantum Matter,
Beijing 100871, China}
\author{Jian-Wei Qiu}
\email{jqiu@bnl.gov}
\affiliation{Theory Center, Jefferson Lab, 12000 Jefferson Avenue, Newport News, VA 23606, USA}
\author{Shinsuke Yoshida}
\email{shinyoshida85@gmail.com}
\affiliation{Theoretical Division, Los Alamos National Laboratory,
Los Alamos, NM 87545, USA}

\date{\today}

\begin{abstract}
Quasi-parton distribution functions have received a lot of attentions in both perturbative QCD and lattice QCD communities in recent years because they not only carry good information on the parton distribution functions, but also could be evaluated by lattice QCD simulations.  However, unlike the parton distribution functions, the quasi-parton distribution functions have perturbative ultraviolet power divergences because they are not defined by twist-2 operators.  In this paper, we identify all sources of ultraviolet divergences for the quasi-parton distribution functions in coordinate-space, and demonstrate that power divergences, as well as all logarithmic divergences can be renormalized multiplicatively to all orders in QCD perturbation theory.
\end{abstract}

\maketitle
\allowdisplaybreaks

\section{Introduction}	
\label{sec:intro}

Parton distribution functions (PDFs), $f_{i/h}(x,\mu^2)$, are defined as the probability distributions to find a quark, an antiquark, or a gluon ($i=q,\bar{q},g$, respectively) in a fast moving hadron $h$ to carry the hadron's momentum fraction between $x$ and $x+dx$,
probed at the factorization scale $\mu$ \cite{Collins:1989gx}. They are fundamental and important nonperturbative functions in QCD quantifying the relation between a hadron and the quarks and gluons within it, and playing an essential role to connect colliding hadron(s) to short-distance QCD dynamics in high energy scattering processes \cite{Brambilla:2014jmp}. However, calculation of PDFs from the first principle, both analytically or from lattice QCD (LQCD), is a challenge, if it is not impossible, due to the fact that PDFs contain the dynamics at long-distance scales and of nonperturbative in nature, and are defined in terms of time-dependent operators.  Traditionally, PDFs have been extracted from high energy scattering data by QCD global analysis in the framework of QCD factorization \cite{Dulat:2015mca,Martin:2009iq,Ball:2014uwa,Alekhin:2013nda,Ethier:2017zbq}.

Recently, Ji introduced a set of quasi-PDFs for a hadron of momentum $p_z$, say along $z$-direction, and argued that they are equal to corresponding PDFs when the hadron momentum $p_z$ goes to infinity \cite{Ji:2013dva}.
Without setting $p_z\to\infty$, the quasi-PDFs could be factorized to PDFs to all orders in QCD perturbation theory so long as quasi-PDFs can be multiplicatively renormalized \cite{Ma:2014jla}.
Like PDFs, the quasi-PDFs are defined by hadronic matrix elements of two-field correlators with a straight line gauge link between the two fields to ensure the gauge invariance,
\begin{eqnarray}
\label{eq:ftq}
\tilde{f}_{q/p}(\tilde{x},\tilde{\mu}^2,p_z)
&\equiv &
\int \frac{d\xi_z}{2\pi} e^{i \tilde{x}p_z \xi_z}\langle h(p)| \overline{\psi}_q(\xi_z)\,\frac{\gamma_z}{2}
\nonumber\\
&\ & \times
\Phi_{n_z}^{(f)}(\{\xi_z,0\})\, \psi_q(0) | h(p) \rangle
\end{eqnarray}
for quasi-quark distribution with $\xi_0=\xi_\perp=0$, and
\begin{eqnarray}
\label{eq:ftg}
\tilde{f}_{g/p}(\tilde{x},\tilde{\mu}^2,p_z)
&\equiv &
\frac{1}{\tilde{x}p_z}
\int \frac{d\xi_z}{2\pi} e^{i \tilde{x} p_z \xi_z}\langle h(p)| F_{z}^{\ \nu}(\xi_z)\,
\nonumber\\
&\ & \times
\Phi_{n_z}^{(a)}(\{\xi_z,0\})\, F_{z\nu}(0) | h(p) \rangle
\end{eqnarray}
for quasi-gluon distribution with $\nu$ summing over transverse directions.
In Eqs.~(\ref{eq:ftq}) and (\ref{eq:ftg}), $\Phi_{n_z}^{(f,a)}(\{\xi_z,0\}) = \text{exp}[-ig\int_0^{\xi_z} d\eta_z\, A^{(f,a)}_z(\eta_z)]$ are the gauge links with ``$f$'' and ``$a$'' representing fundamental and adjoint representation, respectively.
Unlike PDFs, the two-field correlators of the quasi-PDFs are defined to be off the light-cone and have an equal time separation, which makes it possible to calculate the quasi-PDFs in LQCD \cite{Lin:2014zya,Alexandrou:2015rja,Chen:2016utp,Zhang:2017bzy,Orginos:2017kos}.
However, since the hadron momentum in LQCD calculation is effectively bounded by the lattice spacing, the $p_z\to\infty$ limit is hard to achieve in LQCD calculation, and it is a challenge to control the corrections due to the finite $p_z$ \cite{Ji:2014gla,Ji:2017rah}.  Nevertheless, with the potential to calculate the PDFs from the first principle in QCD by using LQCD and the challenges in doing so, the concept of the quasi-PDFs has generated a lot of interests and activities in both perturbative QCD (PQCD) and LQCD community \cite{Ji:2014hxa,Gamberg:2014zwa,Jia:2015pxx,Li:2016amo,Bacchetta:2016zjm,Radyushkin:2016hsy,Radyushkin:2017gjd,Gbedo:2017eyp,Radyushkin:2017ffo,Carlson:2017gpk,Briceno:2017cpo,Nam:2017gzm,Xiong:2017jtn,Radyushkin:2017cyf,Constantinou:2017sej,Yoon:2017qzo,Rossi:2017muf}.  Besides of the corrections due to the limited range of $p_z$, the key to derive PDFs from the LQCD calculated quasi-PDFs is of two-folds:  (1) being able to renormalize all ultraviolet (UV) divergences of quasi-PDFs nonperturbatively, and (2) ensuring the renormalized quasi-PDFs and PDFs to share the same collinear (CO) divergences.

It was demonstrated in Ref.~\cite{Ma:2014jla} by two of the present authors that quasi-PDFs and corresponding PDFs share the same leading logarithmic CO perturbative divergences to all orders in QCD perturbation theory, if the quasi-PDFs can be multiplicatively renormalized.  With this fact, instead of deriving the PDFs from quasi-PDFs by taking the hadron momentum $p_z\to\infty$, two of the present authors proposed in Ref.~\cite{Ma:2014jla} to extract PDFs by using the PQCD factorization approach from LQCD calculated quasi-PDFs or other LQCD calculable single-hadron matrix elements, so long as these matrix elements could be factorized into the desired PDFs with perturbatively calculable coefficients.  In this PQCD factorization approach, the hard scale is of $\tilde{x}p_z \sim 1/\xi_z \sim \text{GeV}$, and the corrections are power suppressed by the hard scale and characterized by the hadronic matrix elements of high twist operators, just like how experimentally measurable and perturbative factorizable hadronic cross sections are connected to PDFs via PQCD factorization.

Therefore, understanding the renormalizability of the operators defining the quasi-PDFs is the most critically important challenge for extracting PDFs from LQCD calculated quasi-PDFs, reliably.  Since the quasi-PDFs are not defined by twist-2 operators, PDFs and quasi-PDFs have different perturbative UV behavior.  Instead of the logarithmic perturbative UV divergence of PDFs, quasi-PDFs have power UV divergences \cite{Ma:2014jla,Xiong:2013bka}.  Although LQCD calculations of quasi-PDFs are naturally regularized by the lattice spacing $a$, the perturbative power divergence in $1/a$ makes it difficult to take the continuous limit of lattice results to extract the correct PDFs \cite{Ishikawa:2016znu,Chen:2016fxx,Monahan:2016bvm,Alexandrou:2017huk, Chen:2017mzz}.  Under the UV renormalization, it is also possible that the operators defining the quasi-PDFs might mix with other operators, for example, the quark PDFs can mix with gluon PDF.  Therefore, it is very important to find out that not only if the operators defining quasi-PDFs are renormalizable, and but also if the operator mixing under the renormalization, if there is any, could be limited to a closed set of operators.  In this paper, we address the issues concerning the renormalizability of the operators defining quasi-PDFs.

The rest of this paper is organized as following.
In Sec.~\ref{sec:def} we will show that quasi-PDFs defined in Eqs.~\eqref{eq:ftq} and \eqref{eq:ftg} have bad short-distance behavior, while coordinate-space quasi-PDFs are better candidates for extracting PDFs. We also define quasi-PDFs in coordinate-space and explain why its renormalization is difficult. We then study the one-loop expansion of quasi-PDFs in coordinate-space in Sec.~\ref{sec:oneloop}. Using the UV power counting derived in Sec.~\ref{sec:powercounting}, we identify all  perturbative UV divergent regions for the quasi-PDFs.  We find that, once subdivergences are subtracted off, all UV divergences are originated from the integration regions where all loop momenta are large, which is significantly different from the behavior of PDFs.
Based these observations and findings, we prove  in Sec.~\ref{sec:renormalization} that quasi-PDFs in coordinate-space can be multiplicatively renormalized. Most importantly, we found that quasi-PDFs do not mix with other quantities under the running of the renormalization scale, which completes our proof of the renormalizability of coordinate-space quasi-PDFs.  Finally, we give our summary in Sec.~\ref{sec:summary}.

\section{Coordinate-space quasi-PDFs}
\label{sec:def}

The interest and excitement of studying quasi-PDFs is based on the potential of extracting PDFs and hadron structure from the first principle calculation of LQCD.  Consequently, the value for studying quasi-PDFs depends on the reliability and accuracy of the matching between quasi-PDFs and PDFs, as proposed in Eq.~(11) of Ref.~\cite{Ji:2013dva},
\begin{equation}
\label{eq:matching_ji}
\tilde{q}(x,\mu^2,P^z) = \int_x^1 \frac{dy}{y} Z\left(\frac{x}{y},\frac{\mu}{P^z}\right) q(y,\mu^2) \, ,
\end{equation}
with $Z(x,\mu/P^z)=\delta(x-1)+(\alpha_s/2\pi)Z^{(1)}(x,\mu/P^z) +\dots \to \delta(x-1)$ as $P^z\to \infty$ for quark distribution.  This is equivalent to require the reliability and accuracy of the PQCD factorization of the quasi-PDFs,
\begin{eqnarray}
\label{eq:factorization}
\tilde{f}_{i/p}(\tilde{x},\tilde{\mu}^2,p_z) & \approx &
\sum_{j} \int_0^1 \frac{dx}{x} {\cal C}_{ij}(\frac{\tilde{x}}{x},\tilde{\mu}^2,\mu^2,p_z)
\nonumber \\
&& \hskip 0.3in
\times f_{j/p}(x,\mu^2) + {\cal O}\left(\frac{1}{\tilde{\mu}^2}\right),
\end{eqnarray}
where $i, j =q, \bar{q}, g$, ${\cal C}$'s are IR safe and perturbatively calculated matching coefficients,  $\tilde{\mu}$ is the renormalization scale of quasi-PDFs, $\mu (\sim{\hskip -0.05in}\tilde{\mu})$ is the factorization scale of PDFs, and the power correction, ${\cal O}(1/\tilde{\mu}^2)$, is characterized by the size of high-twist quark-gluon correlation functions, like in all collinear PQCD factorization formalisms. We found that if the perturbative UV divergences of quasi-PDFs in the left-hand-side (LHS) of Eq.~(\ref{eq:factorization}) are regularized by a momentum cutoff, the factorized convolution in the right-hand-side (RHS) is well behaved. However, the momentum cutoff regulator is hard to implement consistently at higher order calculations in perturbative expansion.  On the other hand, we noticed that if we regularize the perturbative UV divergences of quasi-PDFs by dimensional regularization (DR), the integration over the momentum fraction $x$ on the RHS of the Eq.~\eqref{eq:factorization} will be divergent. This is because sea-quark and gluon PDFs, $f_{j/p}(x,\mu^2) \to x^{-\alpha}$ as $x\to0$, with $1<\alpha<2$, and the calculated matching coefficient, ${\cal C}^{(1)}_{ij}(\tilde{x}/x,\tilde{\mu}^2,\mu^2,p_z)$, evaluated in using DR, has a term proportional to $1/(\tilde{x}/x)$ as $x\to 0$  \cite{Ma:2014jla,Xiong:2013bka}.  As a result, the lower end of the $x$-integration in Eq.~(\ref{eq:factorization}) leads to the divergence, $\int_0 dx/x\, (x/\tilde{x})\, x^{-\alpha} \to \infty$.

By applying the factorization formula in Eq.~(\ref{eq:factorization}) to an asymptotic parton state, ``$j$'', and expanding the both sides of the factorized equation to the first order in power of $\alpha_s$, we have ${\cal C}^{(1)}_{ij}(y,\tilde{\mu}^2,\mu^2,p_z)=\tilde{f}^{(1)}_{i/j}(y,\tilde{\mu}^2,p_z) - f^{(1)}_{i/j}(y,\mu^2)$. As $f^{(1)}_{i/j}(y,\mu^2)$ vanishes as $y\to\infty$, the asymptotic behavior of ${\cal C}^{(1)}_{ij}(\tilde{x}/x,\tilde{\mu}^2,\mu^2,p_z)$ as $x\to 0$ is fully determined by the large $\tilde{x}$ behavior of $\tilde{f}^{(1)}_{i/j}(\tilde{x},\tilde{\mu}^2,p_z)$.

As $\tilde{x}\to\infty$, the large momentum behavior of quasi-PDFs, defined in Eqs.~\eqref{eq:ftq} and \eqref{eq:ftg}, is closely related to the asymptotic behavior of the hadronic matrix elements as the separation of the two fields, $\xi_z\to 0$.  The divergence in the $x$-convolution of the factorized formalism in Eq.~(\ref{eq:factorization}), found above, indicates that hadronic matrix elements could be ill-defined when $\xi_z\to 0$, which is indeed the case as we will demonstrate in the next section.  As pointed out in Ref.~\cite{Ma:2014jla}, the coordinate-space quasi-PDFs are better quantities for the calculation in LQCD, for the discussion of renormalization, and for the extraction of PDFs.

We define coordinate-space quasi-PDFs as following,
\begin{align}\label{eq:ftqx}
\begin{split}
&\tilde{F}_{q/p}(\xi_z,\tilde{\mu}^2,p_z)\\
=& \frac{e^{i p_z \xi_z}}{p_z} \langle h(p)| \overline{\psi}_q(\xi_z)\,\frac{\gamma_z}{2}
\Phi_{n_z}^{(f)}(\{\xi_z,0\})\, \psi_q(0) | h(p) \rangle,
\end{split}
\end{align}
and
\begin{align}\label{eq:ftgx}
\begin{split}
&\tilde{F}_{g/p}(\xi_z,\tilde{\mu}^2,p_z)\\
=&\frac{e^{i p_z \xi_z}}{p_z^2}\langle h(p)| F_{z}^{\ \nu}(\xi_z)\,
\Phi_{n_z}^{(a)}(\{\xi_z,0\})\, F_{z\nu}(0) | h(p) \rangle,
\end{split}
\end{align}
where $\tilde{\mu}$ is a renormalization scale, and $\Phi_{n_z}^{(f,a)}(\{\xi_z,0\})$ are gauge links, defined in Sec.~\ref{sec:intro}. We also define $n_z^\mu = (0,0_\perp, 1)$ with $g^{\mu\nu}=\text{diag}(1,-1,-1,-1)$, and $v\cdot n_z =- v_z$ for any vector $v^\mu$, and we have $n_z^2 = -1$.

We will demonstrate in the next section that coordinate-space quasi-PDFs are well defined for finite $\xi_z$, while they are divergent when $\xi_z\to0$.  if one wants to have a well defined momentum-space quasi-PDFs, one has to properly and consistently subtract off the divergence at $\xi_z\to0$.  From the definitions of quasi-PDFs, we have
\begin{align}
\tilde{F}_{i/p}(-\xi_z,\tilde{\mu}^2,p_z)=\tilde{F}_{i/p}^*(\xi_z,\tilde{\mu}^2,p_z).
\end{align}
To take the advantage of this relation and simplify the discussion, we assume $\xi_z>0$ in the following calculation.  For the final result, we will then express it in the form that is correct for arbitrary value of $\xi_z$.

Before going into the details of proving the renormalizability of quasi-PDFs, we first briefly explain the complexity of quasi-PDFs' renormalization:\\
{\bf 1) Lorentz symmetry is broken.} Because of the explicitly ``$z$"-direction dependence of quasi-PDFs, to identify all possible UV divergences, we need to study both four-dimensional loop momentum integration and three-dimensional loop momentum integration (with ``$z$"-direction fixed)\footnote{Note that these are the only two cases that we need to study. Other cases of loop momentum integration cannot generate new UV divergences.} for each individual loop. This amounts to $2^n$ different cases for a $n$-loop Feynman diagram, which is hard to handle.\\
{\bf 2) Renormalizaiton of composite operators is needed.} As an example, let's choose an $A_z=0$ axial gauge, and the quasi-quark PDFs become
\begin{align}\label{eq:ftqxAz}
\begin{split}
&\tilde{F}_{q/p}(\xi_z,\tilde{\mu}^2,p_z)= \frac{e^{i p_z \xi_z}}{p_z} \left. \langle h(p)| \overline{\psi}_q(\xi_z)\,\frac{\gamma_z}{2}\, \psi_q(0) | h(p) \rangle\right. .
\end{split}
\end{align}
For the renormalization of the quasi-quark PDFs, we need to renormalize the individual quark field, $\overline{\psi}_q(\xi_z)$ and $\psi_q(0)$ in Eq.~\eqref{eq:ftqxAz}, as well as the bi-local composite operator as a whole if it generates UV divergence.  The renormalization of the fields can be naturally taken care of by using the renormalized QCD Lagrangian in $A_z=0$ gauge.  It is the renormalization of the bi-local composite operators that is not covered by the renormalization of QCD.  This is effectively the same procedure for proving the renormalizability of PDFs, for example, for quark PDFs in $A^+=0$ gauge, one has to verify the multiplicative renormalization of the bi-local composite operators defining the quark-PDFs to prove their renormalizability.  It is the renormalization of the bi-local composite operators that mixes quark PDFs with gluon PDF \cite{Collins:1981uw}.  That is, the renormalizability of QCD Lagrangian itself is not enough to guarantee the renormalizability of quasi-PDFs, which are defined by composite operators.

In our explicit calculation and discussion in the rest of this paper, we choose the Feynman gauge since the renormalization of QCD Lagrangian in Feynman gauge is well known.  Since the renormalization of individual quark and gluon field is well defined, we will focus on the renormalization of the composite operators defining the quasi-PDFs.  We will mainly discuss coordinate-space quasi-PDFs, and thus whenever we mention quasi-PDFs,  we mean the coordinate-space quasi-PDFs.  We will first concentrate on quasi-quark PDFs, and then generalize our study to quasi-gluon PDF at the end.

\section{Quasi-quark PDFs at one-loop order}
\label{sec:oneloop}

\begin{figure}[t]
\begin{center}
\includegraphics[width=2.5in]{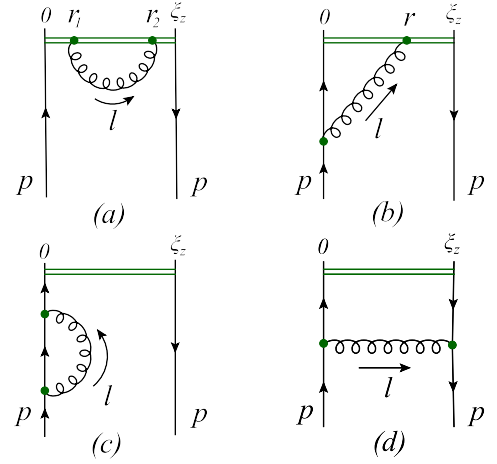}
\caption{\label{fig:oneloop}
Feynman diagrams for quasi-quark PDFs of an asymptotic quark of momentum $p$ at one-loop order.}
\end{center}
\end{figure}

In this section, we study the quasi-quark PDFs at one-loop and its renormalization. Feynman diagrams for quasi-quark PDFs of an asymptotic quark of momentum $p$ at one-loop order in the Feynman gauge are shown in Fig.~\ref{fig:oneloop}.  The diagram (a) in Fig.~\ref{fig:oneloop} gives
\begin{align}
\begin{split}\label{eq:1a0}
M_{1a}=&\frac{e^{i p_z \xi_z}}{p_z}\frac{1}{N_c} \text{Tr}_c[T^aT^a] \int_a^{\xi_z-2a}dr_1\int_{r_1+a}^{\xi_z-a}dr_2\\
&\times \int \frac{d^4l}{(2\pi)^4} \,e^{-i p_z \xi_z} \,e^{il_z(r_2-r_1)} \left(\frac{-ig_{\mu\nu}}{l^2}\right) \\
&\times (-ig_s n_z^\mu)(-ig_s n_z^\nu)  \text{Tr}\left[\frac{1}{2}\, \slashed{p}\, \frac{1}{2}\gamma_z\right]   \\
=& \frac{\alpha_s C_F}{4 i \pi^3} \int_a^{\xi_z-2a}dr_1\int_{r_1+a}^{\xi_z-a}dr_2 \int d^4l\frac{e^{il_z(r_2-r_1)}}{l^2}
\end{split}
\end{align}
where we introduced a cutoff ``$a$'' between fields along the gauge link to regularize potential linear UV divergence\footnote{Although explicit result of UV divergence depends on UV regulators, our conclusions, like power counting rules and renormalization structure, are independent of them.}. The upper limit of $r_1$-integration in Eq. (9) is $\xi_z-2a$ because it needs a separation $a$ from the upper limit of $r_2$-integration.  We will show that this cutoff is enough to regularize all UV divergence in this diagram, and thus we do not introduce DR here. For the following calculation, we introduce a vector $\bar{l}^\mu$, which is the same as $l^\mu$ except that it does not have the $z$-direction component: $l^\mu = \bar{l}^\mu + l_z n_z^\mu$ and $l^2=\bar{l}^2-l_z^2$. We will refer the integration of $\bar{l}^\mu$ as three-dimensional (3-D) integration and the integration of $l^\mu$ as four-dimensional (4-D) integration.  With $d^4l = d^{3}\bar{l}\, dl_z$, let's consider the following phase space integration,
\begin{eqnarray}\label{eq:3dexpansion}
\int \frac{d^{3}\bar{l}}{l^2} &=&
\int \frac{d^{3}\bar{l}}{\bar{l}^2-l_z^2}
\nonumber\\
&=& \int d^{3}\bar{l} \left( \frac{1}{\bar{l}^2}+ \frac{l_z^2}{(\bar{l}^2-l_z^2)\bar{l}^2}\right).
\end{eqnarray}
The first term in Eq.~(\ref{eq:3dexpansion}) is linear divergent, but its coefficient is proportional to $\int dl_z e^{il_z(r_2-r_1)} = 2\pi \delta(r_2-r_1)$, which vanishes because $r_2$ and $r_1$ cannot be at the same point. Thus, effectively, the 3-D integration above is finite. On the other hand, it is easy to see that, if keeping $\bar{l}^\mu$ finite, the integration of $l_z$ is also finite even if $a\to 0$. Therefore, the Eq.~\eqref{eq:1a0} can be UV divergent only in the region where all components of $l^\mu$ go to infinity. As a result, the spacing ``$a$'' can regularize all UV divergences, which gives
\begin{align}\label{eq:1a}
\begin{split}
M_{1a}\overset{\text{div}}{=}-\frac{\alpha_s C_F}{\pi}\frac{|\xi_z|}{a}+\frac{\alpha_s C_F}{\pi}\ln \frac{|\xi_z|}{a},
\end{split}
\end{align}
where we have included the situation when $\xi_z < 0$.

From Fig.~\ref{fig:oneloop}, we have the contribution from the diagram (b) as
\begin{align}
\begin{split}\label{eq:1b0}
M_{1b}=& g_s^2C_F \int_a^{\xi_z-a}dr\int \frac{d^4l}{(2\pi)^4}\,
\frac{e^{il_z r}}{l^2(p-l)^2}
 \\
& \times \frac{1}{p_z} \text{Tr}\left[\frac{1}{2}\, \slashed{p}\, \frac{1}{2}\gamma_z(\slashed{p}-\slashed{l})\gamma_z\right],
\end{split}
\end{align}
where we again use the spacing ``$a$'' as a regulator. For the 3-D integration, it has potential logarithmic divergence from the term proportional to $\slashed{l}$,
\begin{align}
\int \frac{l^\mu\,d^{3}\bar{l}}{l^2(p-l)^2}.
\end{align}
However, the above integration is finite because, to extract the potential logarithmic divergence, we can set the external physical scale $p\to 0$ in the above integration and thus it becomes an odd function in $\bar{l}^\mu$, which vanishes under integration. This explains why the spacing $a$ can regularize the UV divergence of Eq.~\eqref{eq:1b0}. The UV divergence of the diagram (b) is
\begin{align}\label{eq:1b}
\begin{split}
M_{1b}\overset{\text{div}}{=}-\frac{\alpha_s C_F}{2\pi} \ln \frac{|\xi_z|}{a},
\end{split}
\end{align}
where we again included the situation when $\xi_z<0$.

Similarly, we have from the diagram (c) in Fig.~\ref{fig:oneloop},
\begin{align}\label{eq:1c0}
\begin{split}
M_{1c}=&\frac{i g_s^2 \mu_r^{2\epsilon}}{2p_z}(1-\epsilon)C_F\int\frac{d^dl}{(2\pi)^d} \frac{\text{Tr}[\slashed{p}\gamma_z {\slashed{p}}(\slashed{p}-\slashed{l})]}{p^2l^2(p-l)^2}\\
=&i g_s^2 \mu_r^{2\epsilon} C_F(1-\epsilon)\int\frac{d^dl}{(2\pi)^d} \frac{1}{l^2(p-l)^2},
\end{split}
\end{align}
where space-time dimension is defined as $d=4-2\epsilon$. It is clear that the integral in Eq.~(\ref{eq:1c0}) vanishes in DR when $p^2=0$, due to the apparent cancelation between the UV and IR poles in $1/\epsilon$.  The UV divergence from this diagram is logarithmic only if all components of $l^\mu$ go to infinity, and is given in DR by
\begin{align}\label{eq:1c}
\begin{split}
M_{1c}\overset{\text{div}}{=}-\frac{\alpha_s C_F}{4\pi} \frac{1}{\epsilon}.
\end{split}
\end{align}
It is well-known that, at this order and higher orders, UV divergence of massless quark self-energy diagrams can be removed by the renormalization of quark field.

The diagram (d) in Fig.~\ref{fig:oneloop} gives
\begin{align}
\begin{split}\label{eq:1d0}
M_{1d}=& -i g_s^2C_F e^{i p_z\xi_z} \int \frac{d^4k}{(2\pi)^4} e^{-ik_z \xi_z} \frac{1}{k^4 (p-k)^2} \\
&\times \frac{1}{4p_z } \text{Tr}[\slashed{p}\gamma_\mu\slashed{k}\gamma_z\slashed{k}\gamma^\mu]
\end{split}
\end{align}
where $k=p-l$.
Because of the oscillation factor $e^{-ik_z \xi_z}$, contribution from the region where $k_z$ goes to infinity is highly suppressed. Thus, it is UV finite at finite $\xi_z$, although it is divergent as $\xi_z \to 0$,
\begin{align}\label{eq:1d}
\begin{split}
M_{1d}\overset{\text{div}}{=}-\frac{\alpha_s C_F}{2\pi} \ln(|\xi_zp_z|).
\end{split}
\end{align}

In summary, the total one-loop contribution to the UV divergence of quasi-quark PDFs of a quark of momentum $p$ at an arbitrary $\xi_z$ is
\begin{align}\label{eq:1}
\begin{split}
M^{(1)}&\overset{\text{div}}{=}M_{1a}+2\times M_{1b}+ 2\times\frac{1}{2}M_{1c}+M_{1d}\\
&=\frac{\alpha_s C_F}{\pi}\left(-\frac{|\xi_z|}{a} - \frac{1}{4\epsilon} - \frac{1}{2} \ln(|\xi_zp_z|) \right),
\end{split}
\end{align}
where the $ \ln(|\xi_zp_z|)$ term is not UV divergent, but divergent as $\xi_z \to 0$
\footnote{If we take $\xi_z\to 0$, what we calculated is an one-loop correction to a local vector current.  The finite $\xi_z$ in Eq.~\eqref{eq:1} effectively regularizes the UV divergence of the one-loop vertex diagram  Fig. \ref{fig:oneloop}(d), while UV divergence of the self-energy diagram in Fig. \ref{fig:oneloop}(c) is regularized by the dimensional regularization.  From Fig. \ref{fig:oneloop}(d), we obtained the $\ln(|\xi_z p_z|)$ term after we took $\epsilon\to 0$ with $\xi_z$ fixed. If we need to take $\xi_z\to 0$ to mimic the local current, we should keep $\epsilon$ finite to regularize the one-loop UV divergence, which changes $\ln(|\xi_z p_z|)$ to $ - \frac{1}{2\epsilon}$.  As a result, we find from Eq.~\eqref{eq:1} that UV divergences of the vector current vanish at one-loop level if we take $\xi_z\to 0$ with fixed $\epsilon$ and $a$, which is consistent with the expectation of current conservation.}.
 We find that, at this order, UV divergences only come from the region where all loop momenta go to infinity, thus UV divergences are localized in coordinate space.  We will show in the next section that this behavior remains true up to all orders in QCD perturbation theory.

\begin{figure}[ht]
\begin{center}
\includegraphics[width=3.3in]{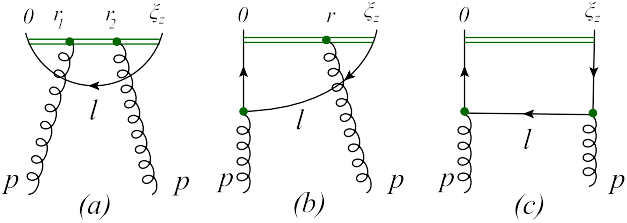}
\caption{\label{fig:oneloopg}
Feynman diagram for quasi-quark PDFs of an asymptotic gluon of momentum $p$ at one-loop order.}
\end{center}
\end{figure}

Feynman diagrams for quasi-quark PDFs of an asymptotic gluon of momentum $p$ at one-loop order in the Feynman gauge are shown in Fig.~\ref{fig:oneloopg}, plus the complex conjugate diagram of (b). A general argument in the next section will show that UV divergences of all diagrams of quasi-PDFs can only come from the 4-D integration, and thus localized in coordinate space. If we keep $\xi_z$ to be finite, all one-loop diagrams in Fig.~\ref{fig:oneloopg} cannot be local, just like the diagram (d) in Fig.~\ref{fig:oneloop}, and therefore, all these one-loop diagrams in Fig.~\ref{fig:oneloopg} must be UV-finite.  But, they can be divergent as $\xi_z\to 0$. To demonstrate this feature, let's take the diagram (a) in Fig.~\ref{fig:oneloopg} as an example. The Fig.~\ref{fig:oneloopg}(a) gives
\begin{align}
\begin{split}\label{eq:2a0}
M_{2a}\propto& \int_0^{\xi_z} dr_1 \int_{r_1}^{\xi_z} dr_2 \int d^4l\, e^{-il_z \xi_z} \frac{l_z}{l^2} \\
=&\frac{\xi_z^2}{2}\int dl_z \, e^{-il_z \xi_z} \, l_z \int d^{3}\bar{l} \left( \frac{1}{\bar{l}^2}+ \frac{l_z^2}{(\bar{l}^2-l_z^2)\bar{l}^2}\right),
\end{split}
\end{align}
which seems to be linearly UV divergent for the 3-D integration. However, similar to the case of diagram (a) in Fig.~\ref{fig:oneloop}, the linear divergent term is proportional to $\delta^\prime(\xi_z)$, which vanishes for finite $\xi_z$. The second term is finite under integration of $\bar{l}$, which gives
\begin{align}
\begin{split}\label{eq:2a}
& \frac{\xi_z^2}{2}\int dl_z \, e^{-il_z \xi_z} \, l_z \int d^{3}\bar{l} \frac{l_z^2}{(\bar{l}^2-l_z^2)\bar{l}^2}\\
\propto&\frac{\xi_z^2}{2}\int dl_z \, e^{-il_z \xi_z} \, \frac{l_z^3}{|l_z|}\\
=&\frac{2i}{\xi_z},
\end{split}
\end{align}
where the proportional relation can be simply obtained by dimensional counting. We therefore find that Fig.~\ref{fig:oneloopg}(a) is UV finite, but behaves as $1/\xi_z$ when $\xi_z \to 0$. Similarly, we found that Figs.~\ref{fig:oneloopg}(b,c) are also UV finite, which indicates that at one-loop, the quasi-quark PDFs get no mixing from a gluon under the UV renormalization.

Before continue, we want to note that the UV behavior found above is significantly different from that of PDFs.  UV divergences of PDFs come from the region of loop momentum $(l_+, l_-, \vec{l}_\perp)\sim(1,\lambda^2,\lambda)$ when $\lambda\to\infty$. Thus, with the momentum component $l_+\sim {\cal O}(1)$, the UV divergence for the normal PDFs is nonlocal in the ``$-$'' direction in coordinate space. It is this fact that the renormalization of PDFs is a convolution, mixed with all twist-2 PDFs with operators along the ``$-$'' direction, while the renormalizaiton of quasi-PDFs is a multiplicative factor as we will show.

We further note that, even if UV divergences ($1/a$, $\ln(a)$ or $1/\epsilon$) are renormalized, the $\ln(|\xi_z|)$ dependence (see Eq.~\ref{eq:1d} as an example) and $1/\xi_z$ dependence (see Eq.~\ref{eq:2a} as an example) of one-loop diagrams signal the divergence of quasi-PDFs as $\xi_z\to 0$, which causes the difficulty in getting a well-defined momentum-space quasi-PDFs if one simply Fourier transforms the coordinate-space quasi-PDFs, as mentioned in the last section. This behaviour will certainly remain true to higher orders in QCD perturbation theory.

\section{Power counting}
\label{sec:powercounting}

\subsection{Superficial UV divergences}

In order to identify all UV divergences from any Feynman diagram contributing to the quasi-PDFs, we start with a set of general diagrams constructed from the lower order Feynman diagrams by adding one more gluon to them. We will then examine how this additional gluon could change the UV divergence of the original diagrams. Specifically, we introduce and study the change of divergence index $\Delta\omega_3$ and $\Delta\omega_4$ corresponding to the 3-D integration and 4-D integration of corresponding loop, respectively. A sufficient condition for quasi-PDFs to be renormalizable is that $\Delta\omega_3\le 0$ and $\Delta\omega_4\le 0$ are satisfied for all cases, but it is not necessary as we will argue. We divide our discussions into five distinctive cases:

\begin{figure}[t]
\begin{center}
\includegraphics[width=3in]{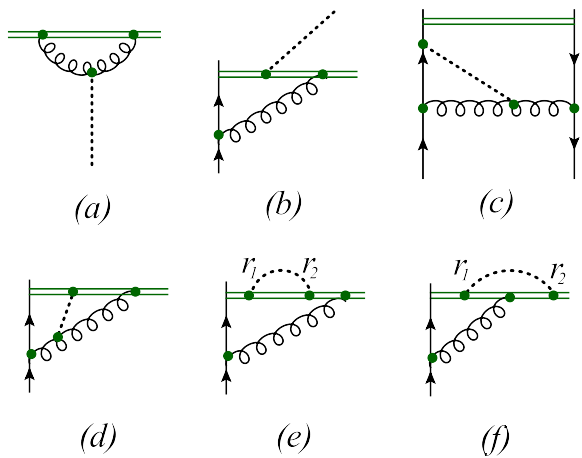}
\caption{\label{fig:addonegluon}
Some loop diagrams, with an additional gluon (denoted as dotted curves) attachments, that could contribute to the quasi-PDFs at higher loop orders.}
\end{center}
\end{figure}

\textbf{Case I:} Only one end of the gluon is attached to parton lines of a loop, like Fig.~\ref{fig:addonegluon}(a). In this case we have one more propagator and one more QCD vertex in the loop, which results in $\Delta\omega_3=-1$ and $\Delta\omega_4=-1$. Thus a linear divergence can be changed to a logarithmic divergence, and a logarithmic divergent diagram is changed to be finite.

\textbf{Case II:} Only one end of the gluon is attached to gauge link of a loop, like Fig.~\ref{fig:addonegluon}(b). In this case, we have one more $z-$direction integration in coordinate space, which does not change the degree of divergence of 3-D integration but reduce the degree of divergence of 4-D integration by 1, such that $\Delta\omega_3=0$ and $\Delta\omega_4=-1$.

\textbf{Case III:} Both ends of the gluon are attached to parton lines of a loop, like Fig.~\ref{fig:addonegluon}(c). In this case we have three more propagators, two more QCD vertexes, and four more momentum integrations for the loop, which results in $\Delta\omega_3=-1$ and $\Delta\omega_4=0$.

\textbf{Case IV:} One end of the gluon is attached to parton lines of a loop, and the other end of the gluon is attached to gauge link of the loop, like Fig.~\ref{fig:addonegluon}(d). In this case we have two more propagators, one more QCD vertex, one more $z-$direction integration in coordinate space, and four more momentum integrations for the loop, which results in $\Delta\omega_3=0$ and $\Delta\omega_4=0$.

\textbf{Case V:} Both ends of the gluon are attached to gauge link of a loop, like Figs.~\ref{fig:addonegluon}(e,f). In this case we have one more propagator, two more $z-$direction integration in coordinate space, and four more momentum integrations for the loop. This results in $\Delta\omega_3= 1$ and $\Delta\omega_4=0$.

A few comments for the above power counting rules are in order.
1) We only concentrated on overall divergences but not on subdivergences, because subdivergences can be taken care by forest subtraction in the step of renormalization. 2) The change of divergence indexes discussed here are only for superficial divergence. It means that even if the power counting indicates that a diagram is divergent, it may not be really divergent due to other considerations. But, on the other hand, if the power counting indicates that a diagram is finite, it must be UV finite.

The \textbf{Case V} is dangerous for renormalization, because $\Delta\omega_3>0$ means that the number of potentially UV divergent topologies is not finite, and thus a finite number of renormalization constants may not be enough to remove all UV divergences. However, as we pointed out, the above power counting rules are only for superficial divergence. We will show in the next subsection that all Feynman diagrams of quasi-PDFs do not have overall UV divergence if only the 3-D integration is considered for any of its loop momenta. As a result, only $\Delta\omega_4$ is relevant for identifying real UV divergent topologies.

\subsection{Finiteness of the 3-D integration}

\begin{figure}[h]
\begin{center}
\includegraphics[width=1in]{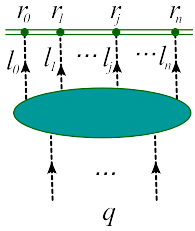}
\caption{\label{fig:connected}
A general ``gauge-link-irreducible" (GLI) diagram, where $l_0=q-l_1-\cdots-l_n$.}
\end{center}
\end{figure}

To demonstrate that the 3-D integration of Feynman diagrams contributing to the quasi-PDFs cannot generate a real UV divergence, we consider the ``gauge-link-irreducible" (GLI) diagrams.  Similar to the concept of one-particle-irreducible (1PI) diagrams, a GLI diagram is a diagram that is still connected even if all gauge links are cut (or removed).  For example, the diagrams (a), (b) and (c) in Fig.~\ref{fig:oneloop} are not the GLI diagrams, while the diagram (d) is.    Similarly, the diagrams (a), (c) and (d) in Fig.~\ref{fig:addonegluon} are the GLI diagrams, while the diagrams (b), (e) and (f) are not.   For studying the renormalization of quasi-PDFs, since the renormalizaiton of QCD Lagrangian is well known, we only need to study the UV properties of the general GLI diagrams as shown in Fig.~\ref{fig:connected}, where the dashed lines connecting to gauge link can be either gluon or quark (if it is at the end point of the gauge link).

In Fig.~\ref{fig:connected}, we assume that the momentum flows into the blob is $q$,  $l_1,\cdots,l_n$ are $n$ loop momenta, and $l_0=q-l_1-\cdots-l_n$ is determined by the momentum conservation. Note that this assignment of momentum is always possible for a GLI diagram. Then,  the general diagram in Fig.~\ref{fig:connected} can be expressed as
\begin{align}\label{eq:connected}
e^{i q_{z} r_0} \prod_{j=1}^n \int_{r_{j-1}+a}^{r_{max}-a}d r_j \int \frac{d^4l_j}{(2\pi)^4} e^{i l_{jz} (r_j-r_0)} \mathcal{M}(q,l_1,\cdots,l_n),
\end{align}
where $\mathcal{M}(q,l_1,\cdots,l_n)$ denotes the blob combined with propagators of the $n+1$ lines connecting to the gauge link.  Since the GLI diagrams can be constructed from one-loop diagrams in Figs.~\ref{fig:oneloop} or~\ref{fig:oneloopg}  combined with insertions in \textbf{Cases I, III} and \textbf{IV} in the last subsection, their overall superficial UV divergence index $\omega\leq1$, no matter we apply the 3-D integration or 4-D integration to each loop momentum.

Now let us study the case in which only the 3-D integration is performed for $l_j$, while carrying out either 3-D or 4-D integration for other loop momenta.  The $\mathcal{M}(q,l_1,\cdots,l_n)$  has two kinds of dependence on $l_j$. One is the polynomial dependence in the numerator, for which we can simply decompose $l_j$ to $\bar{l}_{j}$ and $l_{jz}$. The other is in the propagator like $1/(l_j+k)^2$ where $k$ can be vanishing or depending on other loop momenta. We can decompose the propagator as
\begin{align}\label{eq:decomZ}
\begin{split}
\frac{1}{(l_j+k)^2}=& \frac{1}{\Delta-2k_z l_{jz}-l_{jz}^2}\\
=& \frac{1}{\Delta}+ \frac{2k_z l_{jz}}{\Delta^2}+\frac{(\Delta+4k_z^2+2k_zl_{jz}) l_{jz}^2}{(\Delta-2k_z l_{jz}-l_{jz}^2)\Delta^2},
\end{split}
\end{align}
where $\Delta=(\bar{l}_j+\bar{k})^2-k_z^2$ independent of $l_{jz}$. Based on dimensional counting, each additional $l_{jz}$ in the numerator will suppress the divergence index $\omega$ of the 3-D integration by one unit. Thus the last term in Eq.~\eqref{eq:decomZ} can be safely ignored as we are considering UV divergence from the 3-D integration of $l_j$.  Since the other terms factorize out the $l_{jz}$ dependence from $\mathcal{M}(q,l_1,\cdots,l_n)$, potential UV divergences of these terms are proportional to
\begin{align}
\int d l_{jz} e^{i l_{jz} (r_j-r_0)}  l_z^m \propto \delta^{(m)}(r_j-r_0),
\label{eq:deltam}
\end{align}
with $m$ being non-negative integer. As $r_j$ cannot equal to $r_0$ as defined in Eq.~\eqref{eq:connected}, $\delta^{(m)}(r_j-r_0)$ in Eq.~(\ref{eq:deltam}) vanishes before we take $a\to 0$ limit. Therefore, UV divergences of Fig.~\ref{fig:connected}, obtained by integrating out $\bar{l}_j$ and other loop momenta but fixing $l_{jz}$, eventually vanish after the integration of $l_{jz}$.

In summary, the overall UV divergences of GLI diagrams only come from the region where all loop momenta are large. Furthermore, the third term in Eq.~\eqref{eq:decomZ} is responsible for real UV divergences from the 4-D integration of the GLI diagrams.

\begin{figure}[h]
\begin{center}
\includegraphics[width=1.3in]{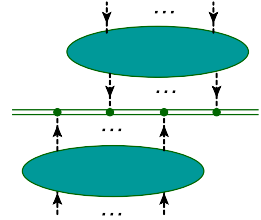}
\caption{\label{fig:disconnect}
A diagram made of two ``gauge-link-irreducible'' (GLI) sub-diagrams.}
\end{center}
\end{figure}

\begin{figure}[h]
\begin{center}
\includegraphics[width=1.1in]{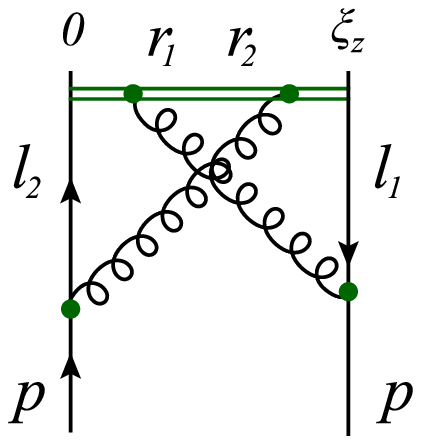}
\caption{\label{fig:twoloop}
A two-loop diagram that might give UV divergent contribution to quasi-quark PDF based on the power counting rules and building blocks in Figs.~\ref{fig:oneloop} and \ref{fig:oneloopg}.}
\end{center}
\end{figure}

Now let us study a diagram made of two GLI sub-diagrams as shown in Fig.~\ref{fig:disconnect}. This non-GLI diagram can be generated either from the diagram in Fig.~\ref{fig:twoloop} or by the insertion of \textbf{Cases II} and \textbf{ V} in the last subsection. Based on the power counting rules, this diagram has overall superficial UV divergence index $\omega\leq2$. When extracting overall UV divergences, we can follow the above discussion for each GLI sub-diagram, and found that the overall UV divergence of the combined diagram vanishes if we fix the $z$-component of any loop momentum. This conclusion can be easily generalized to any diagram made of more GLI sub-diagrams.

We thus conclude that overall UV divergences of any diagram that contributes to quasi-PDFs can only come from the region where all loop momenta are large. An immediate consequence of this finding is that to identify UV divergent topologies, we only need to consider the value of $\Delta\omega_4$ when we are evaluating the five case insertions discussed in the last subsection.  Since we found that $\Delta\omega_4\le 0$ for all cases, there should be only finite number of topologies of UV divergent Feynman diagrams that could contribute to the quasi-PDFs, which we will present in the next subsection.

Let us conclude this subsection by explaining the key difference in UV behavior between quasi-PDFs and PDFs, which have UV divergence from the 3-D integration. As pointed out at the end of Sec.~\ref{sec:oneloop}, UV divergences of PDFs are obtained from a different 3-D integration of loop momentum, say $l_-$ and $\vec{l}_\perp$. We can certainly do a similar decomposition of propagators as that in Eq.~\eqref{eq:decomZ},
\begin{align}\label{eq:decomPlus}
\begin{split}
\frac{1}{(l+k)^2}=& \frac{1}{\hat{\Delta}+2l_+(l_-+k_-)}\\
=& \frac{1}{\hat{\Delta}}- \frac{2(l_-+k_-)l_+}{\hat{\Delta}^2}+\frac{4(l_-+k_-)^2l_+^2}{(\hat{\Delta}+2l_+(l_-+k_-))\hat{\Delta}^2},
\end{split}
\end{align}
where $\hat{\Delta}=2k_+(l_-+k_-)-(\vec{l}_\perp+\vec{k}_\perp)^2$, which is independent of $l_+$. We can also argue that, because $l_+$ is factorized out, the first two terms do not contribute after the integration of $l_+$. However, the last term in Eq.~\eqref{eq:decomPlus} can still generate the UV divergence from the  3-D integration. This is because the UV divergence for PDFs comes from the region where any loop momentum $l$ behaves as $(l_{+}, l_{-}, \vec{l}_{\perp})\sim(1,\lambda^2,\lambda)$ as $\lambda\to\infty$, and thus $l_- l_+ \sim l_\perp^2 \sim \lambda^2$. The boost invariance ensures that each $l_+$ in the numerator will be either accompanied with a ``$-$" momentum component in the numerator or a ``+'' momentum component in the denominator, and neither of these cases suppresses UV divergence index of the loop momentum integration.

\begin{figure}[t]
\begin{center}
\includegraphics[width=3in]{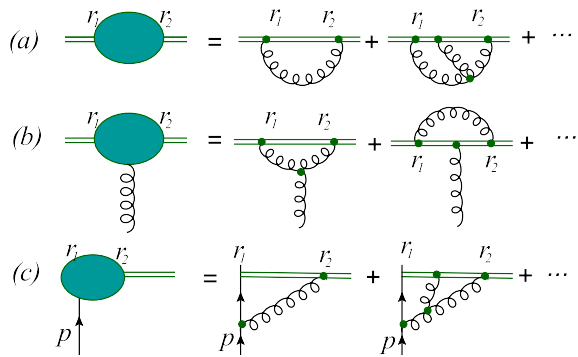}
\caption{\label{fig:multiloopdiv}
Three topologies of diagrams that could give UV divergent contributions to the quasi-quark PDFs.}
\end{center}
\end{figure}

\subsection{Divergent diagrams for quasi-quark PDFs}

Based on the above strategy and power counting rules, we can find out all UV divergent Feynman diagrams begin with a complete set of building blocks. To identify all UV divergent Feynman diagrams for quasi-quark PDFs, we need one more diagram in Fig.~\ref{fig:twoloop} to form the set of building blocks in addition to the one-loop diagrams in Figs.~\ref{fig:oneloop} and~\ref{fig:oneloopg}, because the superficial UV divergence of this two-loop diagram does not agree with that obtained from Fig.~\ref{fig:oneloop}(b) by inserting one more gluon like in \textbf{Case IV}.  Because we only need to consider 4-D integrations, non-vanishing $\xi_z$ in Fig.~\ref{fig:twoloop} ensures that this diagram is UV finite.

Using the above building blocks, we can generate all possible higher order Feynman diagrams. Among them, there are only three types of topologies that could give UV divergent contribution to the quasi-quark PDFs as shown in Fig.~\ref{fig:multiloopdiv}, in addition to those from the renormalizaiton of QCD Lagrangian. From the above discussion, we know that all of these three topologies can be UV divergent only in the region where all loop momenta go to infinity, thus UV divergent contributions are localized in coordinate space. An immediate consequence is that to make the diagrams in Fig.~\ref{fig:multiloopdiv} divergent, $r_2-r_1$ must go to 0. This finding also explains the result of the two-loop calculation in Ref.~\cite{Ji:2015jwa}, where one finds that UV divergence of quasi-quark PDF under DR is proportional to $\delta(1-z)$ in momentum space.

\begin{figure}[h]
\begin{center}
\includegraphics[width=3in]{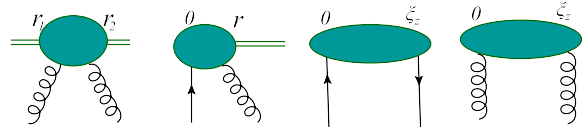}
\caption{\label{fig:multiloopfin}
Sample topologies of diagrams that give UV finite contributions to the quasi-quark PDFs.}
\end{center}
\end{figure}

In addition to these three UV divergent topologies in Fig.~\ref{fig:multiloopdiv},  we also show some topologies that are UV finite in Fig.~\ref{fig:multiloopfin}.  Especially, the last diagram in Fig.~\ref{fig:multiloopfin} indicates that quasi-quark PDFs do not mix with quasi-gluon PDF under the renormalization to all orders in QCD perturbation theory.

\begin{figure}[h]
\begin{center}
\includegraphics[width=3in]{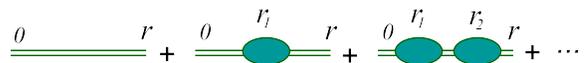}
\caption{\label{fig:sumSE}
Contributions to the general diagrams in Fig.~\ref{fig:multiloopdiv}(a) with all loop diagrams reorganized in terms of 1PI diagrams.}
\end{center}
\end{figure}

\section{Renormalization}
\label{sec:renormalization}

In this section, we present our general arguments for the renormalization of the three topologies of diagrams that could give UV divergent contribution to the quasi-quark PDFs, as identified in the last section.

Let us first consider the power UV divergence from the general diagram in Fig.~\ref{fig:multiloopdiv}(a), which could be expressed in terms of the sum of diagrams made of 1PI diagrams, as shown in Fig.~\ref{fig:sumSE}.  Since the UV divergence is local in coordinate space, we assign the ``i-th'' blob in Fig.~\ref{fig:sumSE} a specific point $r_i$. We express the linear UV power divergence of the ``1st'' blob in Fig.~\ref{fig:sumSE} as $c \int_0^{r} d r_1$, where perturbative coefficient $c=c^{(1)}+c^{(2)}+\cdots$ where $c^{(1)}=-\frac{\alpha_s C_F}{\pi}\frac{1}{a} $ is the first order expansion in $\alpha_s$, as derived in Eq.~(\ref{eq:1a}).

With the geometric sum of the 1PI diagrams, as shown in Fig.~\ref{fig:sumSE}, we can derive all linear UV power divergent contribution to the general diagram in Fig.~\ref{fig:multiloopdiv}(a) by summing over contributions of 1PI diagrams located between 0 and $r\,(r>0)$ to all orders as,
\begin{eqnarray}
&\ & 1+ c \int_0^{r} d r_1 + c^2 \int_0^{r} d r_1 \int_{r_1}^{r} d r_2 + \cdots
\nonumber \\
&\ & \hskip 0.3in
= {\cal P} e^{\, c \int_0^r dr'} = e^{\, c\, r}\, ,
\end{eqnarray}
with ${\cal P}$ indicating the path ordering.
Therefore, one could introduce an overall factor $e^{-c\, |\xi_z|}$ to remove all linear power UV divergences of the quasi-quark PDFs. This overall factor could be thought as the mass renormalization of a test particle moving along the gauge link. Renormalizing power divergence in this way was first proposed in Ref.~\cite{Polyakov:1980ca}.

Besides power UV divergence, there are also logarithmic UV divergences from Fig.~\ref{fig:multiloopdiv}(a).  It is well known~\cite{Dotsenko:1979wb} that these divergences can be removed by a ``wave function'' renormalization of the test particle, $Z_{wq}^{-1}$.

The general diagrams in Fig.~\ref{fig:multiloopdiv}(b) have only the logarithmic UV divergences, effectively from the loop corrections to the gluon coupling to the gauge link.  it was proven in Ref.~\cite{Dotsenko:1979wb} that these logarithmic UV divergences can be absorbed by the coupling constant renormalization of QCD. Therefore, we do not need to worry about them if we use the renormalized QCD Lagrangian.

Finally, let us examine UV divergence from the general diagrams shown in Fig.~\ref{fig:multiloopdiv}(c). Unlike the general diagrams in Figs.~\ref{fig:multiloopdiv}(a) and (b), the loop momentum of diagrams in Fig.~\ref{fig:multiloopdiv}(c) go through active quark (or gluon for the case of quasi-gluon PDF).  Since the UV divergence from the diagrams in Fig.~\ref{fig:multiloopdiv}(c) effectively comes from high order loop corrections to the quark-gauge-link vertex, which is not a fundamental coupling in QCD Lagrangian, using the renormalized QCD Lagrangian to do the calculation does not help remove this kind of UV divergences.  That is, the renormalziation of the operators defining quasi-PDFs is required to remove this kind of UV divergences, if we want to have renormalizable quasi-PDFs.

The key question is then if the operators defining quasi-PDFs will mix with other operators under renormalization? If it does, there could be a danger that the operators defining quasi-PDFs may not form a closed set under the renormalization.

The quark-gauge-link vertex at the lowest order is a simple gamma matrix $\gamma_z$ for the quasi-quark PDFs (is the same for the $A_z=0$ case). As we demonstrated in the last section, UV divergence comes only from the region where all loop momenta are very large, thus we can set $p=0$ if we are only interested in leading UV divergence, which is logarithmic as demonstrated in Sec.~\ref{sec:powercounting}. In addition, the logarithmic UV divergence, $\ln(a)$ as $a\to 0$, is local in coordinate space, and does not have a direct dependence on $\xi_z$.  That is, we find that the UV divergent term of Fig.~\ref{fig:multiloopdiv}(c) only depends on the vector $n_z^\mu$, and consequently, it is proportional to $\gamma_z$ with a constant coefficient, which is proportional to the quark-gauge-link vertex at the lowest order.  Therefore, a constant counterterm is sufficient to remove this kind of UV divergences. Using bookkeeping forests subtraction method, it is straight forward to remove the high order divergences and to show that the net effect is to introduce a constant multiplicative renormalizaton factor $Z_{vq}^{-1}$ for the quark-gauge-link vertex.

In summary, by using renormalized QCD Lagrangian in Feynman gauge, we find that all remained perturbative UV divergences of the quasi-quark PDFs can be removed by introducing a multiplicative renormalization factor, with the multiplicative renormalization factor calculated order by order in QCD perturbation theory.  We expect that similar arguments should also apply for quasi-gluon PDF.  We thus can define the renormalized coordinate-space quasi-PDFs as
\begin{align}\label{eq:renor}
\begin{split}
&\tilde{F}^{R}_{i/p}(\xi_z,\tilde{\mu}^2,p_z)=e^{-C_i |\xi_z|}Z_{wi}^{-1}Z_{vi}^{-1}\tilde{F}^{b}_{i/p}(\xi_z,\tilde{\mu}^2,p_z),
\end{split}
\end{align}
where $C_i$, $Z_{wi}$ and $Z_{vi}$ are renormalization constant depending on parton flavor ``$i$'' but independent of $\xi_z$, and perturbatively calculable order-by-order in powers of $\alpha_s$.   We conclude that the coordinate-space quasi-PDFs are renormalizable and they do not mix with each other or with any other operators under renormalization group equation.

\section{Summary}
\label{sec:summary}

We demonstrated that the behavior of UV divergences of quasi-PDFs is very different from that of PDFs.  While the renormalization of quasi-PDFs is a simple multiplicative factor, the renormalization of PDFs is of a convolution form, due to the fact that the UV divergences of PDFs are not completely local, and consequently, PDFs of different flavors mix with each other under the renormalization.

We show that the locality in space-time of the perturbative UV divergences of the coordinate-space quasi-PDFs makes it possible to have the coordinate-space quasi-PDFs multiplicatively renormalizable, as shown in Eq.~(\ref{eq:renor}), to all orders in QCD perturbation theory.  With the all-order arguments for factorizing all leading power CO divergences of quasi-PDFs into PDFs, given in Ref.~\cite{Ma:2014jla}, we conclude that the renormalized quasi-PDFs could be good candidates for extracting PDFs from LQCD calculations.

For the LQCD calculation of quasi-PDFs, however, the introduction of the overall factor, $e^{-C_i |\xi_z|}$, with perturbatively calculated $C_i$, is not sufficient to remove all power divergences, since we cannot calculate the $C_j$ to all orders.  That is, we need to renormalize the power UV divergence of quasi-PDFs, nonperturbatively.  As shown in Ref.~\cite{Ishikawa:2016znu}, for example, we could introduce an overall non-perturbative factor to remove this kind of power divergences to all orders, for which we will not get into the details here.  In principle, the choice to renormalize the power divergences of quasi-PDFs nonperturbatively is not unique.  But, so long as the renormalization is multiplicative, the renormalization procedure does not alter the CO properties of quasi-PDFs (so that the quasi-PDFs could be factorized into PDFs), the difference in renormalization procedures/choices corresponds to different renormalization schemes, which should lead to different matching coefficients between quasi-PDFs and PDFs.

In addition, we show that the coordinate-space quasi-PDFs are well behaved for all values of $\xi_z$, except when $\xi_z=0$.  That is,
if we want to Fourier transform the coordinate-space quasi-PDFs to derive momentum-space quasi-PDFs, we will have to define a consistent subtraction scheme to remove all divergent terms as $\xi_z\to 0$, before the Fourier transformation.  Without this subtraction, the momentum-space quasi-PDFs are ill-defined at large $\tilde{x}$ region.

{\it Note added:} Two independent studies of the renormalization of quasi-PDFs appeared recently \cite{Ji:2017oey,Green:2017xeu}, in which the same conclusion is reached although  approaches are very different from ours.

\section*{Acknowledgments}

We thank Z.Y. Li and F. Yuan for useful discussions.  This work is supported in part by the Department of Energy, Laboratory Directed Research and Development (LDRD) funding of BNL, under contract DE-SC0012704 (T.I.), the U.S. Department of Energy, Office of Science, Office of Nuclear Physics under Award No.~DE-AC05-06OR23177 (J.Q.), within the framework of the TMD Topical Collaboration (J.Q.), the U.S. Department of Energy, Office of Science under Contract No. DE-AC52-06NA25396 and the LANL LDRD Program (S. Y.), and  JSPS Strategic Young Researcher Overseas Visits Program for Accelerating Brain Circulation (No.R2411) (S.Y.).

\providecommand{\href}[2]{#2}\begingroup\raggedright\endgroup

\end{document}